%% file: hyperbolic-dynamics.tex
\documentclass[journal]{vgtc}

\usepackage{float}

\usepackage[bookmarks,backref=true,linkcolor=black]{hyperref} 
\hypersetup{
  pdfauthor = {},
  pdftitle = {},
  pdfsubject = {},
  pdfkeywords = {},
  colorlinks=true,
  linkcolor= black,
  citecolor= black,
  pageanchor=true,
  urlcolor = black,
  plainpages = false,
  linktocpage
}

\onlineid{326}
\vgtccategory{Research}
\vgtcinsertpkg

\manuscriptnote{}

\usepackage{amsmath,amssymb,amsthm,mathtools,thmtools}

\usepackage{algpseudocode,algorithm}

\usepackage{subfig}

\usepackage{import}

\usepackage{latexsym}
\usepackage{stmaryrd}
\usepackage{subfig}

\usepackage[mathscr]{euscript}

\usepackage{empheq}

\usepackage{listings}
\usepackage[noabbrev,capitalize]{cleveref}
\crefformat{equation}{(#2#1#3)}

\DeclareMathAlphabet{\mathbfsf}{\encodingdefault}{\sfdefault}{bx}{n}

\input macros

\DeclarePairedDelimiter\Parens{\lparen}{\rparen}

\DeclarePairedDelimiter\VBar{\lVert}{\rVert}

\allowdisplaybreaks

\usepackage{newtxtext,newtxmath}
\DeclareMathAlphabet{\mathpzc}{OT1}{pzc}{m}{it}

\author{Andrew McCaleb Reach and Chris North}
\title{Smooth, Efficient, and Interruptible Zooming and Panning}

\authorfooter{
\item
 Andrew McCaleb Reach is with Virginia Tech. E-mail: caleb.reach@vt.edu.
\item
 Chris North is with Virginia Tech. E-mail: north@cs.vt.edu.
}

\shortauthortitle{Reach \MakeLowercase{\textit{et al.}}: Dynamical
  Systems in Hyperbolic Space}

\abstract{This paper introduces a novel technique for smooth and efficient
  zooming and panning based on dynamical systems in hyperbolic space.  Unlike
  the technique of van Wijk and Nuij, the animations produced by our technique
  are smooth at the endpoints and when interrupted by a change of target.  To
  analyze the results of our technique, we introduce world/screen diagrams, a
  novel technique for visualizing zooming and panning animations.}

\begin{document}
\maketitle
\input hyperbolic-dynamics-content
\bibliographystyle{abbrv}
\bibliography{refs.bib}
\end{document}

%% file: macros.tex
\let\V\mathbf

\DeclareMathOperator{\sign}{sign}

\DeclareMathOperator\Log{Log}
\DeclareMathOperator\Exp{Exp}

\DeclareMathOperator\asinh{asinh}
\DeclareMathOperator\acosh{acosh}
\DeclareMathOperator\sech{sech}
\newcommand{\TrMap}{\mathcal T}

\DeclareMathOperator{\Real}{Re}
\DeclareMathOperator{\Imag}{Im}

%% file: hyperbolic-dynamics-content.tex
\section{Introduction}

Zooming and panning is an interaction technique that solves the
problem of fitting too-much data on a too-small screen.
\emph{Zooming} refers to the manipulation of scale---zooming in shows
less in more detail, and zooming out shows more in less detail.
\emph{Panning} refers to the manipulation of the viewing
window---panning left or right slides more information into view from
one side of the screen, while information on the other side slides out
of view.

In many user interfaces, the user zooms and pans manually by pinching
a touch screen, rolling a scroll wheel, or by dragging the mouse.
This paper does not focus on these problems---instead, this paper
introduces techniques for \emph{automatically} zooming and panning in
response to user interactions.

For example, consider a home search application that shows homes
matching the user's query in two ways: in a list and on a map.  A user
who hovers the mouse over a home in the list will likely want to also
see the home's location on the map, and if this home is currently
off-screen, this will require zooming and panning.  If the user
interface performs this navigation abruptly, jumping instantaneously
from the old view to the new view, then the user loses all context
about the spatial relationship between the current house and the
previous house.  By smoothly zooming and panning from one view to the
next, the user gains a clearer understanding of the locations of the
houses.

It can also be useful to combine manual zooming and automatic zooming.
For example, imagine a stock price time series visualization in which
the user manually zooms and pans on the time axis.  As the user
adjusts the time axis view, previously off-screen data will become
visible, and some of this data may be outside the current range of the
price axis.  To prevent the user from having to manually readjust the
price axis after zooming and panning on the time axis, it would be
helpful for the interface to automatically zoom and pan on the price
axis in order to continually accommodate the current range of prices.

The current state-of-the art technique for automatic zooming and
panning is the smooth and efficient technique introduced by van Wijk
and Nuij \cite{van2003smooth}, which seeks to find a navigation
animation that satisfies two properties:
\begin{enumerate}
\item The animation is \emph{smooth} in the second order, i.e. both
  position and velocity are continuous.
\item The animation is \emph{efficient}, i.e. it minimizes perceptual
  cost.  For example, panning directly from a zoomed-in view of
  Seattle to a zoomed-in view of London is perceptually costly because
  the animation will require a long duration in order to avoid
  excessive motion blurring.  By contrast, an animation that first
  zooms out from Seattle, then pans, then zooms into London is
  perceptually cheaper, since this animation can be made significantly
  shorter without excessive motion blurring.
\end{enumerate}
The perceptual cost is quantified in terms of the \emph{optical flow},
and a cost-minimizing zooming and panning animation is found via
methods from differential geometry.

The animations produced by the technique of van Wijk and Nuij are smooth and
efficient while in motion, but they technically violate the smoothness property
at the beginning and end of the animation, where the velocity abruptly jumps
from and to zero respectively.  Moreover, if the animation is interrupted, then
each interruption will, in general, introduce a velocity discontinuity.  Prior
research has shown preliminary evidence that animations without velocity
discontinuities allow users to more easily track objects than animations with
velocity discontinuities \cite{dragicevic2011temporal}.  For the home-finding
example, an interruption occurs every time the user hovers over a different
home, and for the stock price example, interruptions occur continuously while
the user zooms or pans.  This renders the smooth and efficient zooming and
panning technique unsuitable for situations in which animations are frequently
interrupted by a change of target.

We solve the problem of smooth, efficient, and interruptible zooming
and panning by applying ideas from the fields of signal processing and
Riemannian geometry.  Specifically, this paper makes the following
contributions:
\begin{itemize}
\item We introduce the \emph{hyperbolic model of zooming and panning},
  in which a zoom and pan position is represented by a point in the
  Poincar\'e upper half-plane model of hyperbolic space.  The
  hyperbolic model is a simplified version of $u,w$-space.
\item Using this model, we show that smooth, efficient, and
  interruptible zooming and panning can be achieved by generalizing
  signal processing techniques to hyperbolic space.
\item To visualize the results of our technique, we introduce
  \emph{world\slash{}screen diagrams}, a novel method for visualizing
  zooming and panning trajectories that emphasizes the perceptual
  aspects of zooming and panning.
\end{itemize}

\section{Background}\label{sec:background}

Zooming and panning is conceptually simple, but the details can be
difficult to get right, and so several models have been proposed to
aid in the understanding of zooming and panning.  The space-scale
diagrams of Furnas and Bederson \cite{furnas1995space} provide one way
to represent zooming and panning.  In this model, we imagine
\emph{world space} (i.e. the large information space that we are
navigating) is projected from a stationary projector located at the
origin.  To zoom and pan, we maneuver a blank screen through the
space-scale diagram.  By bringing the blank screen closer to the
projector, the projected image appears smaller on the screen, and so
this motion corresponds to zooming out; and by moving the blank screen
farther away from the projector, the projected image appears larger on
the screen, and so this motion corresponds to zooming in.  By moving
the blank screen from side to side, different portions of the
projected image will fall onto the screen, and so this motion
corresponds to panning.  A zooming and panning animation can be
plotted as a screen-trajectory in a space-scale diagram, allowing
zooming and panning animations to be visualized and compared.

\begin{figure}[tb]
  \centering
  \includegraphics{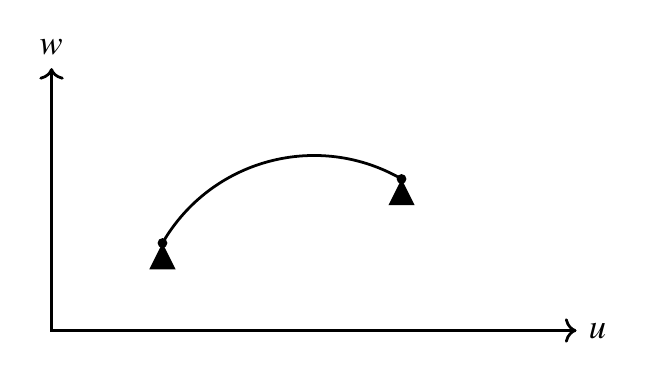}
  \caption{In $u,w$ space, a view is encoded by a point in two-dimensional space
    that represents the location of a camera.  The $u$ coordinate specifies the
    location in world space that is mapped to the center of screen space.  The
    $w$ coordinate specifies the width of the world-space span that is shown
    onscreen.  A zooming and panning animation can be depicted as a path through
    $u,w$ space, as shown.}
  \label{fig:u-w-space}
\end{figure}

Another model of zooming and panning is the $u,w$-space model of van Wijk and
Nuij \cite{van2003smooth,van2004model}.  In the $u,w$-space model, we imagine
that world space is a stationary plane that is viewed through a camera.  To zoom
out, we move the camera away from the world plane, and to zoom in, we move the
camera towards the world plane.  To pan, we move the camera from side to side.
In $u,w$-space, the location of the camera is represented by $u$, which gives
the pan position, and $w$, which gives the zoom factor.  Specifically, $u$ is
the world-space point that is mapped to the center of screen space, and $w$ is
the world-space length spanned by the width of the screen.  The camera has a
angle of view of approximately $53^\circ$.  Additionally, in $u,w$-space, there
is a \emph{metric}, which approximates the perceptual cost of a zooming and
panning animation.  Using this metric, cost-minimizing animation paths were
found by van Wijk and Nuij.  This metric has a parameter $\rho$ that controls
the trade-off between zooming and panning, and when $\rho=1$, this model is
equivalent to hyperbolic geometry \cite{van2004model}.  A diagram depicting
$u,w$ space is shown in \cref{fig:u-w-space}.

Like $u,w$-space, the model we use in this paper is based on a camera
analogy and a perceptual cost metric.  Unlike $u,w$-space, we do not
assume that the camera has a specific angle of view, and in exchange,
our perceptual cost metric is simpler without loss of generality.
Specifically, our metric is always the metric for hyperbolic geometry,
and the trade-off between zooming and panning is instead expressed by
varying the camera's angle of view rather than by varying the metric.

This paper considers methods for automatically zooming and panning between
views.  There is much prior work on manual zooming and panning
\cite{guiard2004view,bederson2000jazz,cockburn2008review} where the user
controls the zoom and pan positions directly.  Prior work on automatic zooming
and panning includes the speed-dependent automatic zooming technique of Igarashi
and Hinckley \cite{igarashi2000speed}, which automatically controls the zoom
position in response to user-performed scrolling.  Specifically, in the
technique of Igarashi and Hinckley, the user controls the rate of scrolling, and
the system automatically adjusts the zoom factor to accommodate the current
scroll rate, i.e. the system zooms out as the scroll rate increases and zooms
back in as the scroll rate decreases.  A technique based on the $u,w$-space
model has also been applied to this problem \cite{van2004model}.  While ideas
presented in this paper may have applications to the problem of speed-dependent
automatic zooming, this paper will only focus on the problem of automatic
zooming and panning between specified views.


\section{Zooming and panning representations}

This section introduces a representation of zooming and panning in
which each specific zooming and panning view is represented by the
location of a camera.  This model is the same as the $u,w$-space model
of van Wijk and Nuij except for one aspect: we do not make any
assumptions about the field of view of the camera.  By not assuming a
particular field of view, we simplify calculations in the following
sections compared with $u,w$-space.  We will consider both the case of
zooming and panning over a two-dimensional space and the case of
zooming and panning over a one-dimensional space.

For zooming and panning in a one-dimensional space, each zoom and pan
position can be described as a point $\V x \in \mathbb H^2$ that represents
the location of a camera, where the set $\mathbb H^2$ is defined by
\[
  \mathbb H^2 = \left\{\begin{bmatrix}u\\v\end{bmatrix}
    :
    u \in \mathbb R\hspace{1pt},\,
    v \in \mathbb R_{>0}
  \right\}
\]
We can decompose a camera position $\V x \in \mathbb H^2$ into two parts as
follows:
\[
  \V x = \begin{bmatrix}u\\v\end{bmatrix}
\]
We will refer to the $u$ component as the \emph{footprint} of the
camera, and to the $v$ component as the \emph{altitude} of the camera.
The camera points straight down at the $u$ axis, and whichever portion
of the $u$ axis is captured by the camera is the portion of
world-space that is shown onscreen.

For zooming and panning in two dimensions, each zoom and pan position
can be represented as a point $\V x \in \mathbb H^3$ that gives the location
of a camera in three-dimensional space, where the set $\mathbb H^3$ is defined
by
\[
  \mathbb H^3 = \left\{\begin{bmatrix}\V u\\v\end{bmatrix}
    :
    \V u \in \mathbb R^2,\,
    v \in \mathbb R_{>0}
  \right\}
\]
As before, we will decompose this point into two parts:
\[
  \V x = \begin{bmatrix}\V u\\v\end{bmatrix}
  = \begin{bmatrix} u_1\\u_2\\v\end{bmatrix}
\]
where $\V u$ is the footprint and $v$ is the altitude.  This camera
points down at the $v=0$ plane.

To avoid needless repetition, we will describe our solution in a
manner that applies at once to both the one-dimensional and
two-dimensional cases.  We will use $\mathbb H^n$ to denote the set of zooming
and panning positions, where $n=2$ for one-dimensional zooming and
panning and $n=3$ for two-dimensional zooming and panning.  The camera
location is therefore a point $\V x \in \mathbb H^{n}$, and the footprint $\V
u \in \mathbb R^{n-1}$ is then either a one-dimensional vector or a
two-dimensional vector.

The relationship between a world-space point $\V p \in \mathbb R^n$
and the corresponding screen-space point $\V r \in \mathbb R^n$ is
given by
\begin{equation}\label{eq:screen-to-world}
  \V p = v\V r + \V u
\end{equation}
This can, of course, be written in inverse form as
\[
  \V r = \frac{\V p - \V u}{v}
\]
which better reflects the more common operation in visualization of
mapping from world space to screen space\footnote{The choice of having
  a simpler mapping from screen space to world space than the other
  way around may seem odd, but it is the price paid for the camera
  analogy, and, more importantly, this choice simplifies later
  equations.}.

The specific cropping of screen space defines the field of view of the
camera.  For a $90^\circ$ camera, screen space is the interval
$[-1,1]$ when $n=1$ and the square region $[-1,1] \times [-1,1]$ when
$n=2$.

\section{The perceptual cost metric}

In this section, we will now introduce a \emph{metric}, i.e. a way to
quantify distance.  Our metric will define distance in the space of
zooming and panning positions to be an approximation of perceptual
cost.

We will quantify the perceptual cost $ds$ incurred by a small change
$d\V x$ in camera position by
\begin{equation}\label{eq:hyper-metric}
  ds = \frac{\VBar{d\V x}}{v}.
\end{equation}
where $\VBar{d\V x}$ is the Euclidean norm of $d\V x$ and $v$ is the altitude of
the camera position $\V x$.  In other words, perceptual cost is proportional to
the Euclidean distance that the camera moves (which includes both changes in
footprint and altitude), and inversely proportional to the camera's altitude.
Notice that unlike Euclidean space, in which distance depends only on the
displacement vector, the metric we have introduced depends on both the camera's
displacement vector $d\V x$ and on the camera's altitude $v$.

The set $\mathbb H^{n}$ together with \cref{eq:hyper-metric} is the
Poincar\'e upper half-space model of hyperbolic space, which has been
widely studied in the field of Riemannian geometry.  We will therefore
frequently refer to camera positions as points in hyperbolic space.

This choice of metric can be explained intuitively.  Imagine a camera
with a $90^\circ$ field of view.  For an altitude of one meter,
panning by half of the screen will require a movement of one meter.
For an altitude of two meters, panning half the screen will require a
movement of two meters.  For an altitude of three meters, panning half
the screen will require a movement of three meters.  Clearly then, the
onscreen effect of panning depends on both the distance that the
camera moves and the altitude of the camera.  For a given altitude,
moving the camera a greater distance will, of course, result in a
greater onscreen movement: therefore perceptual cost should be
proportional to the distance the camera moves.  However, since a
proportionally larger distance is required for the same perceptual
cost at higher altitudes, the perceptual cost of panning should be
inversely related to altitude.

Now consider zooming.  At an altitude of one meter, the camera must
move upwards (i.e. in the direction of increasing altitude) by one
meter to zoom out by a factor of two.  We assume that zooming out by a
constant factor will have the same perceptual cost regardless of where
the camera is located.  At an altitude of two meters, the camera must
move upwards by two meters to zoom out by a factor of two.  At an
altitude of three meters, the camera must move upwards by three meters
to zoom out by a factor of two.  Therefore, like panning, the
perceptual cost of zooming depends on both the distance traveled and
the altitude, and like panning, the cost of zooming should be
proportional to the distance zoomed and inversely proportional to
altitude\footnote{It might seem as though the relationship between
  camera distance and the perceptual cost of zooming is more
  complicated than this, since we would expect zooming out by a factor
  of two to have the same perceptual cost as zooming in by a factor of
  two, but at an altitude of two meters a two-meter movement is
  required to zoom out by a factor of two, whereas only a one-meter
  movement is required to zoom in by a factor of two.  However, this
  discrepancy becomes smaller with shorter movements of the camera,
  and since the metric quantifies the effect of an infinitesimal
  movement of the camera, this discrepancy disappears.}.

The perceptual cost of panning is related to the camera's angle of
view---the distance required to pan one screen for a camera with a
narrow angle of view is much smaller than that required to pan one
screen for a camera with a wide angle of view.  The perceptual cost of
zooming, however, is not related to the camera's angle of
view---zooming out by a factor of two requires the same camera
movement regardless of the angle of view.  Therefore, by varying the
camera's angle of view, different trade-offs can be made between
zooming and panning.  A narrow angle of view expresses a preference
for panning rather than zooming, and a wide angle of view expresses a
preference for zooming rather than panning.

\subsection{Comparison with $u,w$-space}

In the $u,w$-space model, the camera's field of view is fixed, and a parameter
$\rho$ that appears in the metric controls the trade-off between zooming and
panning.  By contrast, in the hyperbolic model, the metric has no parameter to
control the trade-off between zooming and panning---instead the camera has a
variable field of view.  By increasing the field of view, the same trajectory
appears to zoom more and pan less, and by decreasing the field of view, the
trajectory appears to pan more and zoom less.  This is because the same onscreen
view will map to different camera locations depending on the angle of view---a
camera with a narrow angle of view will need to be located at a higher altitude
to capture the same image as camera with a wide angle of view.  In Smooth and
Efficient Zooming and Panning \cite{van2003smooth}, a user study found that the
user-preferred value of $\rho$ was 1.42, which is roughly $\sqrt{2}$.
Interestingly, for $\rho = \sqrt{2}$ in the $u,w$-space model, the corresponding
angle of view in the hyperbolic model is exactly $90^\circ$, as is shown later
in this section.

Our model has the following advantages over $u,w$-space:
\begin{enumerate}
\item The hyperbolic model of zooming and panning is precisely the
  same as the Poincar\'e upper half-space model of hyperbolic space,
  which has been well-studied in Riemannian geometry.  Therefore, a
  large body of work in differential geometry on the half-space model
  immediately applies to zooming and panning.

\item In $u,w$-space, the parameter $\rho$ appears in many of the
  equations that define the optimal zooming and panning trajectory.
  These equations are simpler in the hyperbolic model.

\item In the coordinates of $u,w$-space, the optimal path is an
  elliptical trajectory, where the eccentricity of the ellipse is
  determined by the parameter $\rho$.  As we will see, in the
  coordinates of the hyperbolic model, the optimal path is always a
  simpler circular path.

\item In $u,w$-space, the $w$ parameter is less intuitive if the
  screen is not square, since it gives the length of only one of the
  screen's sides, and the other must be calculated from the screen's
  aspect ratio.  By contrast, in the hyperbolic model, we simple use
  two angles of view instead of one.
\end{enumerate}

The hyperbolic model and $u,w$-space are related by a change of
coordinates.  Under this change of coordinates, the models are
\emph{equivalent up to scaling}, by which we mean that the optimal
paths are the same with respect to this change of coordinates and the
distances between points are related by a constant factor.  We will
now describe more precisely what it means for two models to be
equivalent up to scaling.  One way to interpret the condition of
equivalence up to scaling is that the two models agree in every way
except that distances are calculated in different units, e.g. one
model might give the distance in inches whereas the other model gives
the distance in centimeters.  In practice this difference is
unimportant, since even if a parameter was added to scale the metric
by an arbitrary factor, this parameter would be redundant with other
existing parameters that control how the optimal path is used.  For
example, in the model of van Wijk and Nuij, there is a parameter $V$
that controls how quickly the optimal path is traversed.  If the
metric is scaled by a factor of two, and so all points are twice as
far away from each other, then an animation that matches the original
can be produced by similarly scaling $V$ by a factor of two, therefore
traversing the path at twice the speed and reaching the destination at
the same time as traversing the original path at the original speed.
Therefore, when we say that two models are equivalent up to scaling,
we mean that they are equivalent in all useful ways.

Specifically, the hyperbolic model is equivalent up to scaling to the
$u,w$-space model using the change of coordinates\footnote{In the case
  of $\rho = \sqrt{2}$, which was the user-preferred value as found by
  van Wijk and Nuij, this formula becomes $w = 2v$, which is easily
  remembered since the character $w$ visually appears to be the
  concatenation of two $v$ characters.}
\[
  w = \rho^2 v.
\]
Hyperbolic distance $s$ is related to $u,w$-space distance $\sigma$ by
\[
  s = \rho^2\sigma.
\]
The angle of view $\theta$ in the hyperbolic model is related to
$\rho$ in $u,w$-space by the equation
\[
  \theta = 2 \arctan \frac{\rho^2}{2}.
\]
As mentioned earlier, for $\rho = \sqrt{2}$, as was preferred by users in the
study of van Wijk and Nuij \cite{van2003smooth}, the above equation gives
$\theta = 90^\circ$.

In the appendix to Smooth and Efficient Zooming and Panning
\cite{van2003smooth}, an embedding of $u,w$-space in three-dimensional
space is shown.  In hyperbolic geometry, the well-known
\emph{pseudosphere} gives such an embedding, and can be written in
terms of the footprint $u$ and altitude $v$ as:
\begin{align*}
  x &= (\cos u)/v\\
  y &= (\sin u)/v\\
  z &= \acosh v - \tanh(\acosh v).
\end{align*}
Here $\tanh$ denotes the hyperbolic tangent and $\acosh$ denotes the inverse
hyperbolic cosine.  It should be noted that, like the embedding of van Wijk and
Nuij, this is only a \emph{partial} embedding---in particular, it does not work
for $v < 1$, since $\acosh v$ is imaginary for $v < 1$.  Moreover, this mapping
is ambiguous, since a point $(x,y,z)$ corresponds to multiple values of $u$.


\section{Hyperbolic space}

In this section we will explore the properties of hyperbolic space and
how they relate to zooming and panning.  Fist, we will discuss the
geodesics of hyperbolic space, which correspond to perceptually
optimal zooming and panning paths.  Second, we will discuss vectors in
hyperbolic space and operations that relate vectors and geodesics.  We
will use the properties and operations from this section later in this
paper to develop algorithms for smooth, efficient, and interruptible
zooming and panning.

We begin by exploring the geodesics of hyperbolic space.  It is known
that every geodesic in the Poincar\'e upper half-plane model is either
a half-circle whose endpoints lie at altitude zero or a vertical
line~\cite{cannon1997hyperbolic}.

\subsection{Geodesics and distance}

We will define the function
\[
  \operatorname{geo}(\V x, \V y, s)
\]
to be optimal path between $\V x \in \mathbb H^{n}$ and $\V y \in \mathbb H^{n}$,
parameterized by arc length $s$ as measured by the hyperbolic metric
\cref{eq:hyper-metric}.

\begin{figure}[tb]
  \centering
  \includegraphics{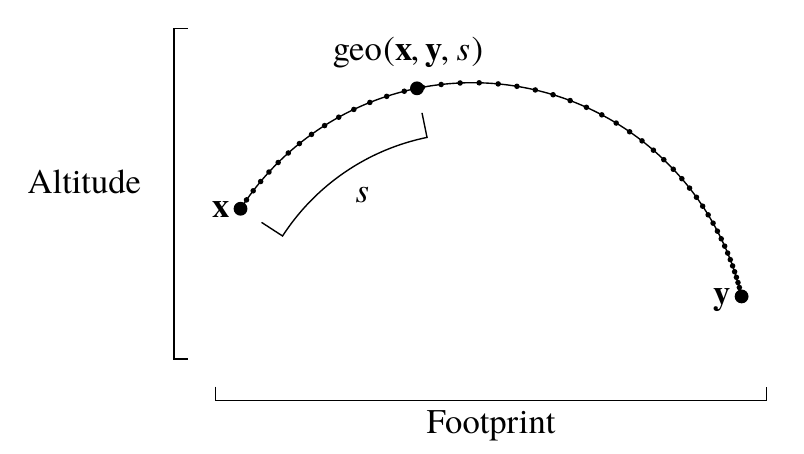}
  \caption[Hyperbolic geodesics]{The function $\operatorname{geo}(\V x, \V y, s)$ returns
    the point arrived at by traveling from $\V x$ along a geodesic to
    $\V y$ for a hyperbolic distance $s$.}
  \label{fig:geodesic-footprint-altitude}
\end{figure}

We will also define the function
\[
  \operatorname{dist}(\V x, \V y)
\]
to be the hyperbolic distance between two camera locations; i.e. the
perceptual cost of the optimal path between these two points.  This
function is the hyperbolic analog of Euclidean distance.  We have the
identities
\begin{align*}
  \operatorname{geo}\Parens*{\V x, \V y, 0} &= \V x\\
  \operatorname{geo}\Parens*{\V x, \V y,
  \operatorname{dist}(\V x, \V y)} &= \V y\\
  \operatorname{dist}\Parens*{\V x,
    \operatorname{geo}\Parens*{\V x, \V y, s}} &= s
\end{align*}
The function $\operatorname{geo}$ is depicted in
\cref{fig:geodesic-footprint-altitude}.

Finally, we will define a third function, $\operatorname{gerp}$, in
terms of $\operatorname{geo}$ and $\operatorname{dist}$ as
\begin{equation}
  \operatorname{gerp}(\V x, \V y, \alpha) =
  \operatorname{geo}\Parens[\Big]{\V x, \V y, \alpha\operatorname{dist}(\V x, \V y)}
\end{equation}
This function interpolates between $\V x$ at $\alpha = 0$ and $\V y$
at $\alpha = 1$ along a perceptually optimal path.



The functions $\operatorname{geo}$ and $\operatorname{dist}$ can be
calculated as
\begin{subequations}
  \begin{align}
    \operatorname{dist}\Parens*{
      \left[\begin{smallmatrix}\V u_0\\v_0\end{smallmatrix}\right],
      \left[\begin{smallmatrix}\V u_1\\v_1\end{smallmatrix}\right]} &= S\\
    \operatorname{geo}\Parens*{
      \left[\begin{smallmatrix}\V u_0\\v_0\end{smallmatrix}\right],
      \left[\begin{smallmatrix}\V u_1\\v_1\end{smallmatrix}\right],
      s} &= \left[\begin{smallmatrix}\V u(s)\\v(s)\end{smallmatrix}\right]
  \end{align}
  where, for $\V u_0 \neq \V u_1$,
  \begin{align}
    \V u(s) &= \V u_0 +
              v_0\frac{\sinh(s)}{\cosh(s + r_0)}
              \frac{\V u_1 - \V u_0}{\VBar{\V u_1 - \V u_0}}
              \label{eq:u-hyper}\\
    v(s) &= v_0\frac{\cosh(r_0)}{\cosh(s + r_0)}\label{eq:v-hyper}\\
    S &= r_1 - r_0\label{eq:s-hyper}\\
    r_i &= \asinh \frac
    {v_1^2 - v_0^2 + (-1)^i \VBar{\V u_1 - \V u_0}^2}
    {-2v_i\VBar{\V u_1 - \V u_0}}\label{eq:r-hyper}
  \end{align}
  and, for $\V u_0 = \V u_1$,
  \begin{align}
    \V u(s) &= \V u_0\label{eq:u-hyper-v}\\
    v(s) &= v_0 e^{sk}\label{eq:w-hyper-v}\\
    k &= \sign(v_1 - v_0)\\
    S &= |\log(v_1/v_0)|.\label{eq:s-hyper-v}
  \end{align}
\end{subequations}
In these equations, $\sinh$ and $\cosh$ denote the hyperbolic sine and cosine
functions respectively, and $\asinh$ denotes the inverse hyperbolic sine.  The
notation $\sign$ denotes the sign function, which returns $-1$ when its argument
is negative, $0$ when its argument is zero, and $1$ when its argument is
positive.  Finally, the notation $\log$ is used to denote the natural logarithm.
These equations were adapted from those given by van Wijk and Nuij
\cite{van2003smooth}, but were reformulated to decrease numerical error by
removing sums and differences that caused catastrophic cancellation.
\Cref{fig:hylerp-streamlines} visualizes the path produced by the function
$\operatorname{geo}(\V x, \V y, s)$ between two views $\V x$ and $\V y$ as a
world\slash{}screen diagram.



\subsection{Vectors in hyperbolic space}

The functions defined in the last subsection operated on points in
hyperbolic space.  In the following subsections, we will define
functions that manipulate both points and \emph{vectors} in hyperbolic
space.  In the same way that a point in hyperbolic space can be used
to represent the \emph{position} of a camera, a vector in hyperbolic
space can be used to represent the \emph{velocity} of a camera.  In
this subsection we will introduce a notion of vectors in hyperbolic
space.  These ideas can be found in most textbooks on differential and
Riemannian geometry
\cite{do1992riemannian,lee2012introduction,lee1997riemannian}.

In Euclidean space, a vector has a direction and a magnitude.  In
Riemannian geometry, a vector additionally has a point at which it's
based.  This is necessary because the fundamental properties of the
space change depending on where the base point is located.  We will
use bold capital letters, e.g. $\V X$, to denote vectors in hyperbolic
space, and we will use the notation $\mathbb T_{\V x}\mathbb H^n$ to
denote the set of all vectors located at point $\V x$.

The hyperbolic magnitude $|\V X|$ of a vector $\V X \in \mathbb T_{\V
  x}\mathbb H^n$ is defined as
\[
  |\V X| = \frac{\|\V X\|}{v}
\]
where $\|\V X\|$ is the Euclidean norm of $\V X$, and $v$ is the
altitude of $\V x$.  This definition mirrors the definition of the
metric.  This must be the case because the magnitude of a velocity
vector must gives a notion of speed, i.e. the rate of change of
distance, and this notion must match the notion of length defined by
the metric.

\subsection{Exponential map}

\begin{figure}[tb]
  \centering
  \includegraphics{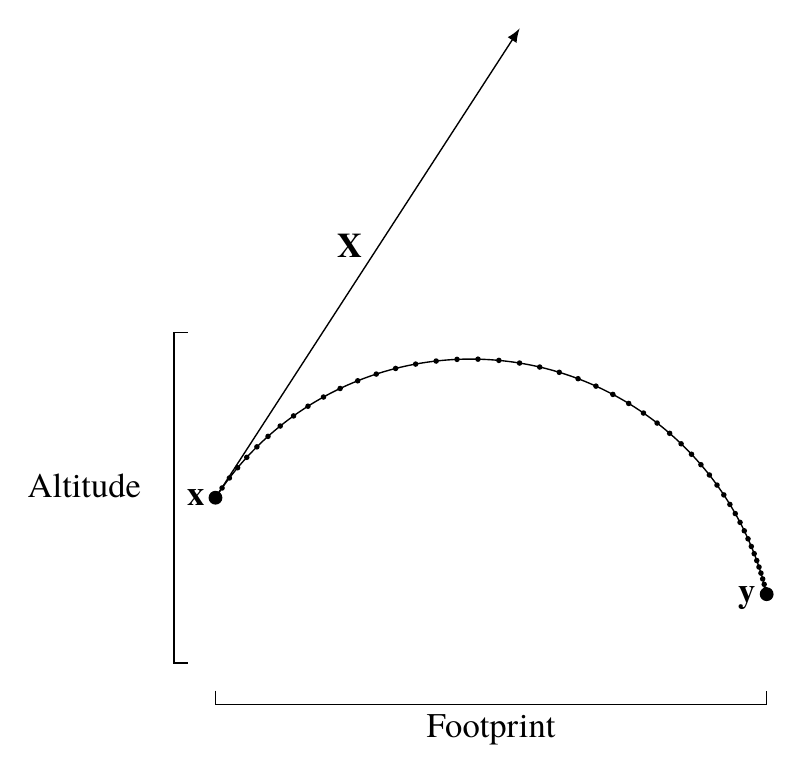}
  \caption[Hyperbolic exponential and logarithmic maps]{The hyperbolic
    exponential map $\Exp(\V X)$ sends a vector $\V X \in \mathbb
    T_{\V x}\mathbb H^n$ to the point $\V y = \Exp(\V X)$ by finding a
    geodesic $\boldsymbol\upgamma(t)$ whose initial velocity
    $\boldsymbol\upgamma'(0)$ is equal to $\V X$ and then calculating
    $\V y = \boldsymbol\upgamma(1)$.  The hyperbolic logarithmic map
    $\Log_{\V x}(\V y)$, by contrast, sends two points $\V x$ and $\V
    y$ to the vector $X = \Log_{\V x}(\V y)$ by finding a geodesic
    $\boldsymbol\upgamma(t)$ whose endpoints are
    $\boldsymbol\upgamma(0) = \V x$ and $\boldsymbol\upgamma(1) = \V
    y$ and then calculating $\V X = \boldsymbol\upgamma'(0)$, where
    $\boldsymbol\upgamma'(0)$ is the initial velocity of
    $\boldsymbol\upgamma(t)$.  The exponential and logarithmic maps
    are therefore inverses of each other.}
  \label{fig:exp-log-maps}
\end{figure}

We will now define a function, the \emph{exponential map}, that takes
a vector in hyperbolic space as input.  The exponential map $\Exp(\V
X)$ of the vector $\V X \in \mathbb T_{\V x}\mathbb H^n$ gives the point arrived at by
traveling a hyperbolic distance $|\V X|$ from $\V x$ along an optimal
path with initial direction matching $\V X$.  This process is depicted
in \cref{fig:exp-log-maps}.

For a vector $\V X \in \mathbb T_{\V x}\mathbb H^n$ with footprint and altitude
components $\V U$ and $V$ respectively located at the base point $\V
x$ with footprint $\V u_0$ and altitude $v_0$, the exponential map can
be calculated as
\begin{subequations}
\begin{equation}
  \Exp(\V X)
  =
  \left[\begin{smallmatrix}\V u_1\\v_1\end{smallmatrix}\right]
\end{equation}
where, for $\V U \neq 0$,
\begin{align}
  \V u_1 &= \V u_0 +
  \frac{v_0\sinh |\V X|}{\cosh(|\V X| + r_0)} \frac{\V U}{\VBar{\V U}}\\
  v_1 &= v_0\frac{\cosh(r_0)}{\cosh(|\V X| + r_0)}\\
  r_0 &= -\asinh \frac{V}{\|\V U\|}.
\end{align}
and, for $u_0 = u_1$,
\begin{align}
  U &= 0\\
  W &= v_0 \log(v_1/v_0)
\end{align}
\end{subequations}

\subsection{Logarithmic map}

The logarithmic map takes two points $\V x, \V y \in \mathbb H^n$ as arguments
and produces a vector $\Log_{\V x}(\V y)$ that points in the direction
of $\V y$ along an optimal path from $\V x$, where the magnitude of
this vector is equal to $\operatorname{dist}(\V x, \V y)$.  The
logarithmic map can be defined by
\[
  \Log_{\V x}(\V y) = \operatorname{gerp}'(\V x, \V y, 0)
\]
where $\operatorname{gerp}'$ denotes the derivative of
$\operatorname{gerp}$ with respect to its last parameter.  Then
$\Log_{\V x} \V y \in \mathbb T_{\V x}\mathbb H^{n}$ is the initial velocity in hyperbolic
space of the optimal path from $\V x$ at $t=0$ to $\V y$ at $t=1$, and
so this operation is analogous to the subtraction $\V y - \V x$ in
Euclidean space.

Calculations reveal that for $\V x_0, \V x_1 \in \mathbb H$ with respective
footprints $\V u_0, \V u_1$ and altitudes $v_0, v_1$, the logarithmic
map is given by
\begin{subequations}
  \begin{equation}
  \Log_{\V x_0}(\V x_1) = \V X
\end{equation}
where, for $\V u_0 \neq \V u_1$, the footprint $\V U$ and altitude $V$
components of $\V X$ are given by
  \begin{align}
  U &= v_0 S \sech(r_0) \frac{\V u_1 - \V u_0}{\VBar{\V u_1 - \V u_0}}\\
  W &= -v_0 S \tanh r_0\\
  S &= r_1 - r_0\\
  r_i &= \asinh \frac
  {v_1^2 - v_0^2 + (-1)^i \VBar{\V u_1 - \V u_0}^2}
  {-2v_i\VBar{\V u_1 - \V u_0}}
  \end{align}
  If $\V u_0 = \V u_1$, then $(\V U,W)$ is given by
\begin{align}
  U &= 0\\
  W &= v_0 \log(v_1/v_0)
\end{align}
\end{subequations}

The exponential and logarithmic maps are inverses of each other,
i.e.
\[
  \Exp \Parens*{\Log_{\V x} (\V y)} = \V y.
\]
A useful identity that we will use later relating the exponential map,
logarithmic map, and geodesic interpolation is
\begin{equation}\label{eq:gerp-exp-log}
  \Exp\left( t\Log_{\V x}(\V y) \right) =
  \operatorname{gerp}(\V x, \V y, t)
\end{equation}

\subsection{Covariant derivative}

In Euclidean space, we can calculate acceleration by differentiating
velocity.  However, the second derivative of $\operatorname{gerp}(\V
x, \V y, t)$ will generally be nonzero, since geodesics in hyperbolic
geometry are generally curved.  However, geodesics are the hyperbolic
analog of straight lines in Euclidean space, and so its reasonable to
desire a hyperbolic analog of acceleration for which geodesics have
zero hyperbolic acceleration.

In Riemannian geometry, this notion is provided by the \emph{covariant
  derivative}, which provides a way to differentiate time-varying
vectors.  Hyperbolic acceleration is then given by the covariant
derivative of velocity.  The covariant derivative can be defined in
terms of the metric \cite{do1992riemannian}, and calculations reveal
that the covariant derivative for hyperbolic space is given by
\begin{subequations}
\begin{equation}
  \frac{D\V X}{dt} = \V X'
\end{equation}
where the footprint $\V U'$ and altitude $V'$ components of $\V X'$
are given by
\begin{align}
  \V U'
  &=
    \frac{d\V U}{dt}
    -
    \frac{1}{v}
    \Parens*{
    \frac{dv}{dt}\V U
    + \frac{d\V u}{dt} V
    }
    \label{eq:cov-der-u}\\
  V'
  &=
    \frac{dV}{dt}
    +
    \frac{1}{v}
    \Parens*{
    \frac{d\V u}{dt} \cdot \V U - \frac{dv}{dt}V
    }
    \label{eq:cov-der-w}
\end{align}
\end{subequations}
in which $\V u$ and $v$ are the footprint and altitude of $\V x$, the
base point of $\V X$.

\subsection{Transport map}

In Euclidean space, vectors do not need to keep track of their base
points, because a vector formed at one point can be easily used at any
other point.  For example, consider the expression $\V z + (\V y - \V
x)$.  In this expression, the subtraction $\V y - \V x$ is used to
form a vector from base point $\V x$ pointing to $\V y$, and then this
vector is used in an operation with a different point $\V z$.  In
hyperbolic space, things are not so simple.  If we try to directly
change the base of a vector, e.g. change $\V X \in \mathbb T_{\V x}\mathbb H^n$ to $\V Y
\in \mathbb T_{\V y}\mathbb H^n$, then basic properties of the vector may change, e.g. the
hyperbolic magnitude of these two vectors will not match if the
altitude of $\V x$ does not equal the altitude of $\V y$.

\begin{figure}[tb]
  \centering
  \includegraphics{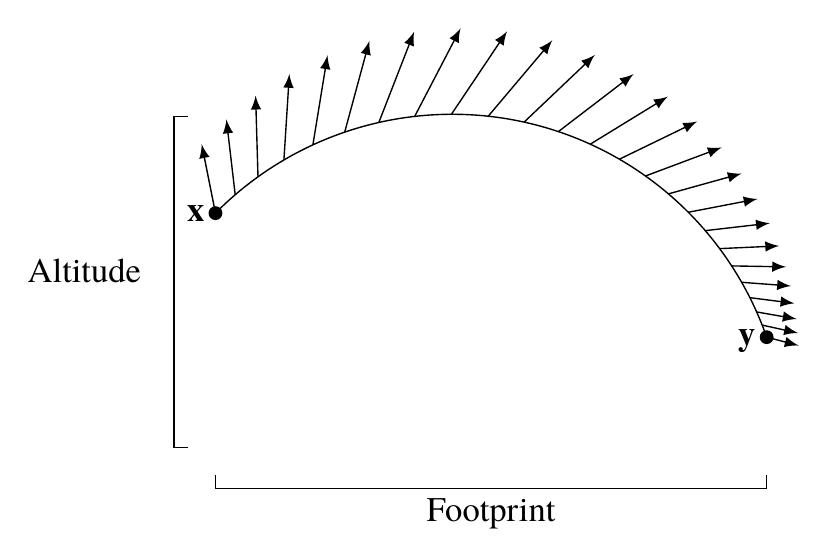}
  \caption[Hyperbolic transport map]{The transport map $\TrMap_{\V y}(\V X)$ of the hyperbolic
    vector $\V X \in \mathbb T_{\V x}\mathbb H^n$ is the vector $\V Y \in \mathbb T_{\V y}\mathbb H^n$
    obtained by parallel transporting $\V X$ to the new base point $\V
    y \in \mathbb H^n$ along a geodesic from $\V x$ to $\V y$.  In hyperbolic
    space, the transport map causes the Euclidean magnitude to change
    so that the hyperbolic magnitude remains the same---if the target
    altitude is higher than the starting altitude, then the Euclidean
    magnitude will grow, and if the target magnitude is lower than the
    starting altitude, then the Euclidean magnitude will shrink.  The
    vector will also rotate, keeping a constant Euclidean angle with
    the geodesic.}
  \label{fig:hyperbolic-transport-map}
\end{figure}

In Riemannian geometry, the \emph{parallel transport} operation allows
a vector to be moved to another location.  The result of parallel
transporting a vector depends not just on the end points and the
vector, but also on the specific path taken.  We will now define a
\emph{transport map}, which moves vectors to another location along a
geodesic.

\begin{subequations}
  For points $\V x_0, \V x_1$ with respective footprints $\V u_0, \V
  u_1$ and altitudes $v_0, v_1$, the transport map of a vector $\V X
  \in \mathbb T_{\V x_0}\mathbb H^n$ to $\V x_1$ is given by
  \begin{equation}
    \mathcal T_{\V x_1}(\V X) =
  \left[\begin{smallmatrix}\V U_1\\V_1\end{smallmatrix}\right]
  \end{equation}
  where, for $\V u_0 \neq \V u_1$,
  \begin{align}
  \V U_1 &= \frac{v_1}{v_0}\V U_0^\perp + \frac{v_1}{v_0}\Real(z_1)\frac{\V u_1 - \V u_0}{\VBar{\V u_1 - \V u_0}}\\
  V_1 &= \frac{v_1}{v_0}\Imag(z_1)\\
  z_1 &= \vartheta_1\overline{\vartheta_0}z_0\\
  z_0 &= U_0^\parallel + W_0\boldsymbol i\\
  U_0^\parallel &= \V U_0 \cdot \frac{\V u_1 - \V u_0}{\VBar{\V u_1 - \V u_0}}\\
  \V U_0^\perp &= \V U_0 - U_0^\parallel\frac{\V u_1 - \V u_0}{\VBar{\V u_1 - \V u_0}}\\
  \vartheta_i &= \tanh r_i + \boldsymbol i\sech r_i\\
  r_i &= \asinh \frac
        {v_1^2 - v_0^2 + (-1)^i \VBar{\V u_1 - \V u_0}^2}
        {-2v_i\VBar{\V u_1 - \V u_0}}
\end{align}
and, for $\V u_0 = \V u_1$,
\begin{align}
  \V U_1 &= \frac{v_1}{v_0}U_0\\
  V_1 &= \frac{v_1}{v_0}V_0\label{eq:hyper-wt-par}
\end{align}
\end{subequations}
In these equations, $\boldsymbol i$ denotes the imaginary unit, and
$\overline{\vartheta_0}$ denotes the complex conjugate of $\vartheta_0$.  The
functions $\Real$ and $\Imag$ the real and imaginary parts of their respective
arguments.  Imaginary numbers are used in these equations to perform rotations.
The result of the transport map is shown in \cref{fig:hyperbolic-transport-map}.

\section{Visualizing zooming and panning animations}

A simple way to visualize a zooming and panning path in $\mathbb H^n$ is to
simply plot the footprint on the $x$ axis and the altitude on the $y$
axis, i.e. simply plot the path the camera takes.  This corresponds to
the $u,w$-space diagrams of van Wijk and Nuij.  However, it can be
difficult to discern the perceptual effect of zooming and panning from
these footprint\slash{}altitude diagrams.

\begin{figure}[tb]
  \centering
  \includegraphics{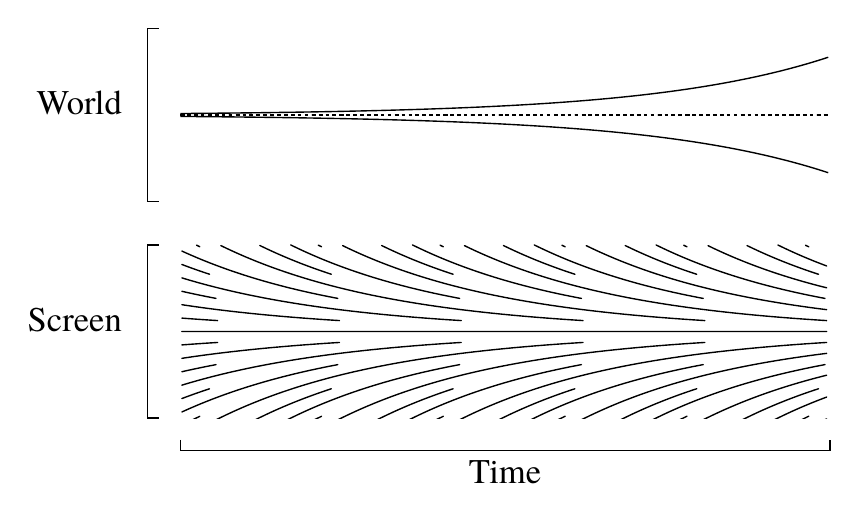}
  \caption[Zoom-out world/screen diagram]{A \emph{world\slash{}screen diagram}
    for a zooming-out animation.  The upper plot, the \emph{screen bounds
      diagram}, shows the screen center and screen bounds plotted in world space
    over time.  The lower plot, the \emph{optical pathline diagram}, shows world
    points plotted in screen space over time.  The slope of each line in the
    optical pathline diagram at each instant in time corresponds to the velocity
    of an onscreen point at that instant.  As the animation progresses, the
    screen occuppies an increasingly large portion of world-space, as can be
    seen in the screen bounds diagram.  At the same time, a given portion of
    world space occupies an increasingly small portion screen space, as can be
    seen from the optical pathline diagram.}
  \label{fig:zoom-out-streamlines}
\end{figure}

\begin{figure}[tb]
  \centering
  \includegraphics{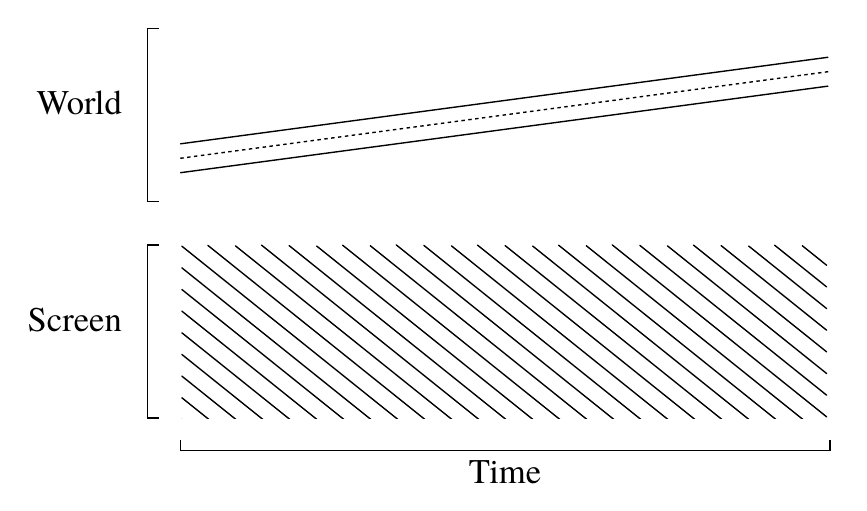}
  \caption[Zoom-out world/screen diagram]{A world\slash{}screen diagram for a
    panning animation.  The world bounds diagram (above) shows that the screen moves
    through world space at a constant rate, and the optical pathline diagram (below)
    shows that world points move through screen space at a constant rate.}
  \label{fig:pan-streamlines}
\end{figure}

\begin{figure}[tb]
  \centering
  \includegraphics{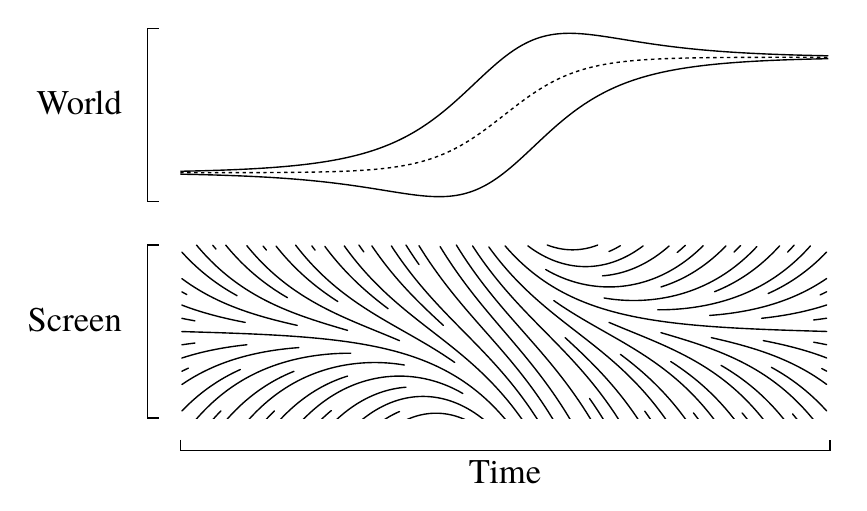}
  \caption[World/screen diagrams]{A world\slash{}screen diagram for optimal
    zooming and panning.  This diagram depicts zooming out, followed by panning,
    followed by zooming in.  The specific zooming and panning path used was
    calculated using the smooth and efficient technique of van Wijk and Nuij.}
  \label{fig:hylerp-streamlines}
\end{figure}

To visualize a zooming and panning animation in a way that emphasizes the
perceived motion, we introduce a novel visualization technique:
\emph{world\slash{}screen diagrams}.  A world\slash{}screen diagram consists of
two sub-diagrams that share a common time axis.  The upper part of the
world\slash{}screen diagram is a \emph{screen bounds diagram}, which plots the
screen center and screen bounds in world space over time.  The lower part of the
world\slash{}screen diagram is an \emph{optical pathline diagram}, which plots
world points in screen space over time.  An optical pathline diagram has time on
the horizontal axis and screen space on the vertical axis, and each curve in the
diagram is a constant point in world space.  For each world-point curve in the
optical pathline diagram, the slope of the curve corresponds to the onscreen
velocity of the world point.  Example world\slash{}screen diagrams are shown in
\cref{fig:zoom-out-streamlines,fig:pan-streamlines,fig:hylerp-streamlines}.  By
mapping time to a dedicated axis, the world\slash{}screen allows for precise
visualizations of the temporal aspects of a zooming and panning animation.  By
contrast, in a $u,w$-space diagram, time must be encoded by color or by dots
placed at regularly-spaced intervals in time.

One advantage of world\slash{}screen diagrams over other
representations of zooming and panning trajectories is that the
optical pathline diagrams corresponds more closely to the perception
of zooming and panning than other diagrams.  In fact, the
root-mean-square optical flow velocity, which van Wijk and Nuij
originally used to define their metric, can be determined for each
time instant in the world\slash{}screen diagram by finding the
root-mean-square slope of all pathlines at that instant.  By contrast,
points taken from a perceptually uniform trajectory at equally-spaced
moments in time will appear to be unevenly spaced if plotted in a
footprint\slash{}altitude diagram or a space-scale diagram, and so
these diagrams do not accurately represent perception.

\subsection{Optical pathline diagrams}

We will now describe the algorithm we used to create the optical
pathline diagrams in this paper.  In principle, an optical pathline
diagram can be created by taking any set of world-space points and
plotting their paths in an optical pathline diagram.  However, this
approach leads to diagrams in which curves are closely spacing in some
regions widely spaced in others.  To remedy this problem, we add and
remove world-point curves from the diagram as space allows.

We will now show that the problem of adding and removing world-point
curves can be seen as a problem of adding and removing isolines
(i.e. contour lines) from a scalar field.  Consider the scalar field
$\varphi$ that maps time $t$ and screen point $r$ to the corresponding
world point $p = \varphi(t,r) = v(t)r + u(t)$.  The isolines of this
field are the world-point curves in the optical pathline diagram.
Therefore, we must choose when to add and remove isolines from the
diagram.

Note that if $\varphi(t,r)$ is linear, then the gradient magnitude
$\VBar{\nabla \varphi(t,r)}$ is constant over all $t$ and $r$.  In this
case, if we add isolines at equally-spaced wold-points, then the
isolines will also be equally spaced.  The wold-point
spacing $P$ and screen-space isoline spacing $R$ are related by
\[
  P = R\VBar{\nabla \varphi(t,r)}
\]
In practice, the function $\varphi(t,r)$ will not be linear, but provided
that it is sufficiently smooth, the above relationship will
approximately hold.  This relationship provides the intuition behind
the algorithm we will describe.

Let the set $C_\alpha \subset \mathbb R^2$, for parameter $\alpha > 0$
that controls the spacing, be the set of all points in the optical
pathline diagram at which a contour should be drawn.  We define
$C_\alpha$ as the set of all $(t,r)$ such that there exists a power of
two not less than $\alpha\VBar{\nabla \varphi(t,r)}$ that divides $\varphi(t,r)$.
Using the notation
\[
  \llceil x \rrceil = 2^{\left\lceil \log_2(x) \right\rceil}
\]
the definition of $C_\alpha$ can be written as
\[
  C_\alpha = \left\{
    (t \in \mathbb R, r \in \mathbb R)
    : \llceil \alpha\VBar{\nabla \varphi(t,r)} \rrceil
    \mathrel{\text{divides}} \varphi(t,r)
  \right\}.
\]
This can be seen as a discretized version of the dense isocontour
imaging technique of Matvienko and Kr\"uger \cite{matvienko2013dense}.

Now that we have defined $C_\alpha$, we will describe an algorithm for
computing the optical pathline diagrams.  The optical pathline diagram
will take as input the parameter $\alpha$ which controls the spacing
of the pathlines, the footprint and altitude functions $u(t)$ and
$v(t)$, and the screen-space bounds $r_{\text{\tiny low}}$ and
$r_{\text{\tiny high}}$.

Notice that
\begin{align*}
  \VBar{\nabla \varphi(t,r)}^2
  &=
  \left(
    \frac{\partial}{\partial t} p(t,r)
  \right)^2 +
  \left(
    \frac{\partial}{\partial r} p(t,r)
  \right)^2\\
  &=
  \left(
    rv'(t) + u'(t)
  \right)^2
  +
  \left(
    v(t)
  \right)^2
\end{align*}
where $v'(t)$ and $u'(t)$ are the derivatives of $v(t)$ and $u(t)$
respectively.  Because
\[
  \left|v(t)\right| = \left|\frac{\partial}{\partial r} p(t,r)\right| \leq \VBar{\nabla p(t,r)},
\]
a necessary condition for $(t,r) \in C_\alpha$ is that $\llceil\alpha\lvert
v(t)\rvert\rrceil$ divides $p(t,r)$.  Taking advantage of this necessary
condition, we can use the following algorithm to find the optical
pathline diagram:
\begin{itemize}
\item Let $A$ be an empty hash map.
\item For each time step $i$:
  \begin{itemize}
  \item Calculate
    \begin{align*}
      P &= \llceil\alpha\lvert v(t_i)\rvert\rrceil\\
      p_{\text{\tiny low}} &= P\left\lceil \frac{v(t_i)r_{\text{\tiny low}} + u(t_i)}{P} \right\rceil\\
      p_{\text{\tiny high}} &= P\left\lfloor \frac{v(t_i)r_{\text{\tiny high}} + u(t_i)}{P} \right\rfloor.
    \end{align*}
  \item For each $p$ in $p_{\text{\tiny low}},\,p_{\text{\tiny low}} +
    P,\,p_{\text{\tiny low}} + 2P,\,p_{\text{\tiny low}} +
    3P,\,\dots,\,p_{\text{\tiny high}}$:
    \begin{itemize}
    \item Calculate
      \[
        \psi = \VBar{\nabla \varphi(t,r)}
        = \sqrt{
          \Parens[\Big]{
            v(t)
          }^2 +
          \Parens[\Big]{
            r v'(t_i) + u'(t_i)
          }^2}
        \]
        If the derivatives $v'$ and $u'$ are not available
        (e.g. because the zooming and panning animation is computed
        frame-by-frame rather than as a continuous path), then they
        can be approximated by finite differences.
      \item If $\llceil \alpha\psi \rrceil$ does not divide $p$ then
        continue to the next $p$.
    \item Otherwise, check to see if $A[p]$ exists.
      \begin{itemize}
      \item If $A[p]$ does not exist, assign to $A[p]$ a list containing
        $i$ as its sole element.
      \item If $A[p]$ does exist, append $i$ to the list $A[p]$.
      \end{itemize}
    \end{itemize}
  \end{itemize}
\item For each key $p$ in $A$:
  \begin{itemize}
  \item For each $i$ in $A[p]$:
    \begin{itemize}
    \item Calculate the screen-space point $r$ corresponding to $p$ using
      \[
        r = \frac{p - u(t_i)}{v(t_i)}.
      \]
      \item If $i$ is the first element in $A[p]$, or if the previous
        element in $A[p]$ was not $i-1$, then begin a new pathline at
        $(t_i, r)$.
      \item Otherwise, add a new segment to the current pathline that
        connects to $(t_i, r)$.
    \end{itemize}
  \end{itemize}
\end{itemize}

\section{The problem of interruptions}

The previous sections have developed the hyperbolic model of zooming and
panning.  We will now use this model to solve the interruptible zooming and
panning problem.

In the interruptible zooming and panning problem, we are given as
input a \emph{target signal} $\V x(t)$ and we must compute as output a
\emph{smoothed signal} $\V y(t)$.  At each instant $t$, the target
camera position is given by $\V x(t) \in \mathbb H^n$, and the smoothed camera
position $\V y(t) \in \mathbb H^n$ must be computed.  Importantly, the
computation of the smoothed camera position $\V y(t)$ must depend only
on $\V x(\tau)$ for all $\tau < t$.  In other words, the
transformation from the target signal to the smoothed signal must be
\emph{causal}---the smoothed camera position must be computed at each
instant using only the past and present behavior of the input signal.
This models the real-world usage of interruptible zooming and panning
in user interfaces, where future user input is unknowable.

\begin{figure*}[t!]
  \subfloat[Constant speed]{%
    \includegraphics{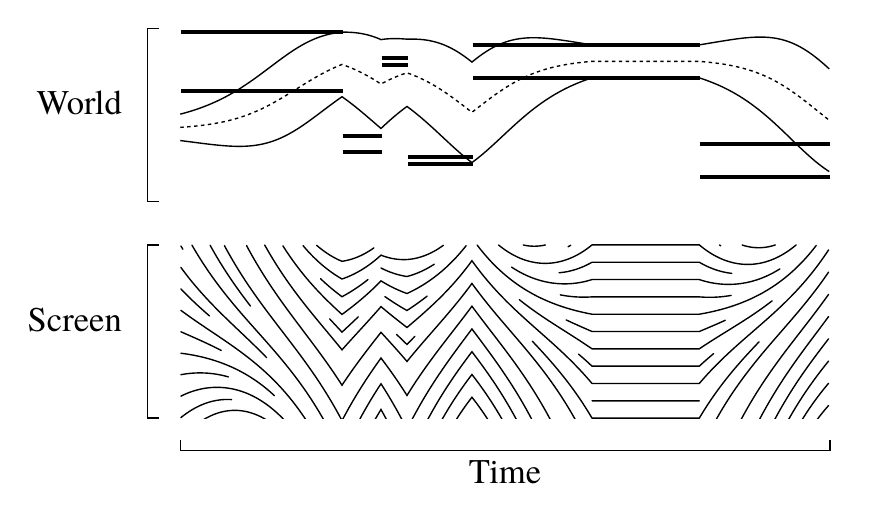}%
    \label{fig:interrupted-hylerp}}
  \subfloat[Easing curve]{%
    \includegraphics{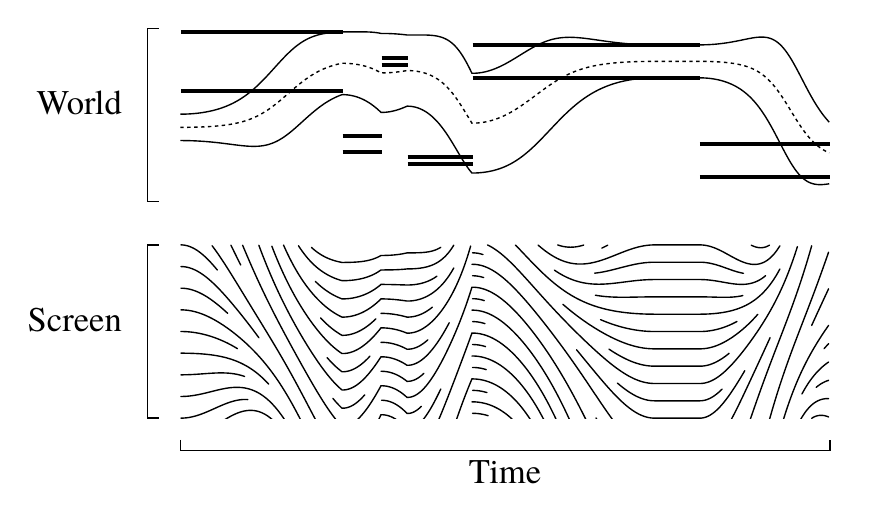}%
    \label{fig:interrupted-easing-hylerp}}\\
  \subfloat[Cascaded one-pole filter]{%
    \includegraphics{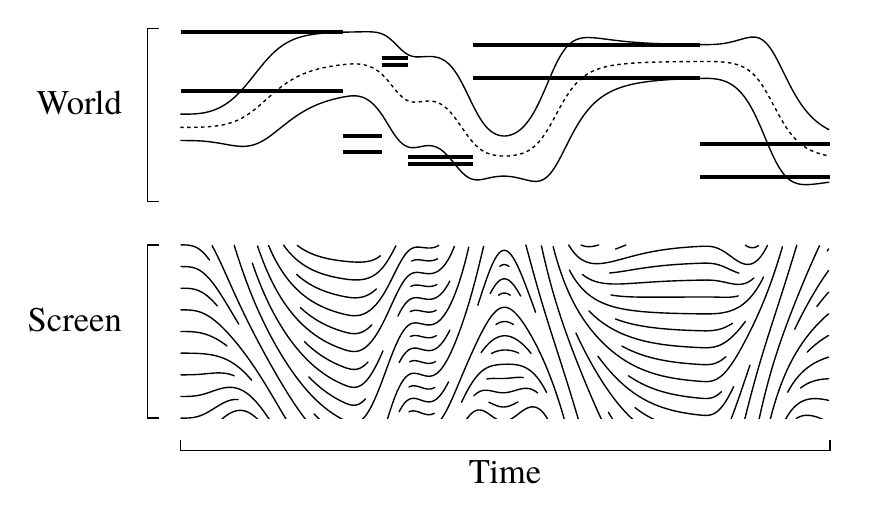}%
    \label{fig:cascaded-one-pole-hylerp}}
  \subfloat[Clipped cascaded one-pole filter]{%
    \includegraphics{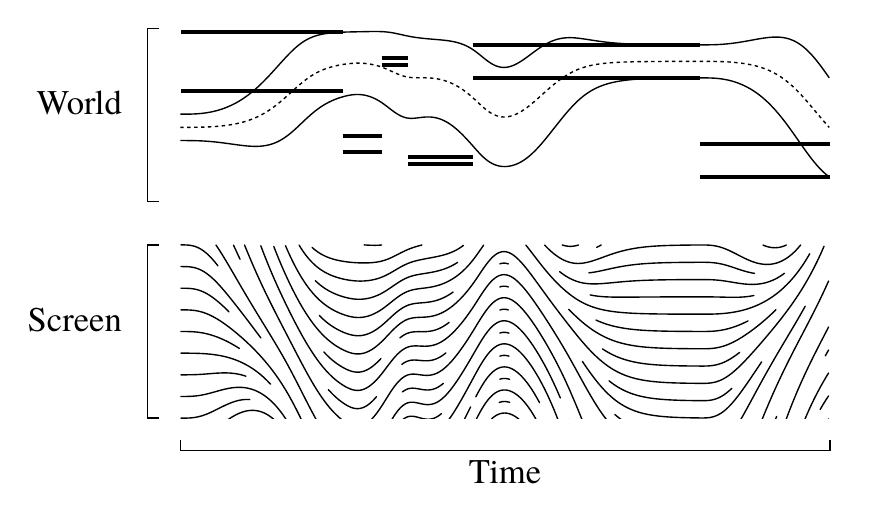}%
    \label{fig:clipped-four-pole-hylerp}}
  \caption[Zooming and panning approaches compared]{These world\slash{}screen
    diagrams depict the result of different zooming and panning approaches
    applied to the same target signal. In these diagrams the target signal is
    depicted as a pair of bold lines that shows the desired world span.  The
    constant-speed technique uses the optimal navigation approach of van Wijk
    and Nuij \cite{van2003smooth} to produce an animation towards the current
    target.  The easing curve technique shows an straightforward modification of
    optimal navigation.  As can be seen, neither the constant-speed technique
    (a) nor the easing curve technique (b) are smooth when interrupted.  For
    both of these techniques, interruptions produce velocity discontinuities in
    screen space.  By contrast, a geodesic one-pole filter cascaded four times
    (c) is smooth even when interrupted, as is the clipped cascaded one-pole
    filter (d).}
\end{figure*}

Before we introduce the signal-processing-based solutions that are the
main focus of this paper, we first describe two simple techniques and
analyze their disadvantages.  These simpler techniques assume that the
target signal is a step function, i.e. the target camera position
changes abruptly at times $t_1,t_2,t_3,\dots$ but is otherwise
constant.

\subsection{Constant speed}

The obvious way to handle an interruption is to simply begin a new
path from the current position to the new target.  With this approach,
the output always moves towards the target at a constant hyperbolic
speed.  Specifically, for each change of target, follow the path
$\operatorname{geo}(\V y_0, \V x_1, ct)$, where $\V y_0$ is the
current output at the time of the target change, $\V x_1$ is the new
target, $c$ is the speed of the animation, and $t$ is the time elapsed
since the target change.  When $ct \geq \operatorname{dist}(\V y_0, \V
x_1)$, the target has been reached, and so the output should remain at
the target until the next change of target.

The problem with this technique is that every time the target is
reached, and every time the target is changed\footnote{Except, of
  course, for the edge case where the new target happens to lie on the
  same geodesic as the current target, and in the same direction.}, a
velocity discontinuity is introduced into the smoothed signal.  These
discontinuities are easily seen in a world\slash{screen} diagram, as
illustrated in \cref{fig:interrupted-hylerp}.

\subsection{Easing curves}

Another simple solution is to use an \emph{easing curve}.  As we will
show, this solves the problem of velocity discontinuities at the
endpoints, but does not solve the problem of interruptions.

The easing curve can be specified as follows: for each change of
target, follow the path $\operatorname{gerp}(\V y_0, \V x_1, f(t/d))$,
where $\V y_0$ is the current output at the time of the target change,
$\V x_1$ is the new target, $d$ is the desired duration of the
animation, $t$ is the elapsed time since the target change, and $f$ is
an \emph{easing curve}, which satisfies $f(0) = 0$ and $f(\alpha) = 1$
for all $\alpha \geq 1$.  The easing curve should also satisfy $f'(0)
= f'(1) = 0$, where $f'$ denotes the derivative of $f$.  One easing
curve that satisfies these properties is
\[
  f(\alpha) = \begin{cases}
    (1/2) - (1/2)\cos(\pi \alpha) & \alpha < 1\\
    1 & \alpha \geq 1,
    \end{cases}
\]
but there are many other options as well.

Provided that $f'(0) = f'(1) = 0$, this approach does not suffer from
velocity discontinuities at the endpoints.  However, it does not
handle the case of interruptions, since an interruption would cause
the velocity to abruptly jump to zero as the new animation segment
begins.  The velocity discontinuities introduced by interruptions in
the easing approach are easily seen in the world\slash{}screen
diagram, as shown in \cref{fig:interrupted-easing-hylerp}.

One difference between this technique and the previous technique is
that the duration of the animations produced using this technique is
fixed, whereas for the previous technique, the duration of the
animations produced is proportional to the distance between the start
point and the target point.  We do not claim that having a fixed
duration is superior to a variable duration or vice versa---the choice
depends on the specific effect the interaction designer is trying to
achieve.


\section{Smoothing using geodesic one-pole filters}

To solve the problem of smoothing both the endpoints and interruptions, we
introduce a novel technique for smooth and efficient zooming and panning that
draws inspiration from signal processing.  While signal processing techniques
have previously been used to animate Euclidean attributes
\cite{reach2017signals}, to process orientation signals \cite{lee2002general},
and to process signals on a sphere \cite{osborne2013geodesic}, to our knowledge
signal-processing techniques have not been applied to signals in hyperbolic
space.

The techniques of the previous section were formulated in terms of
target changes, and each target change triggered a change in the
formulation of the output.  In this section, we take a different
approach.  Instead of processing target change events, we instead
process the target signal directly, transforming the target signal
into the smoothed signal using differential equations.  There are many
ways to perform this transformation, and different methods will
produce different animations.  This section introduces a particularly
simple method, which is based on the \emph{one-pole low-pass filter},
a common filter in signal processing.  We call our technique the
\emph{geodesic one-pole filter}.

\begin{figure}
  \centering
  \includegraphics{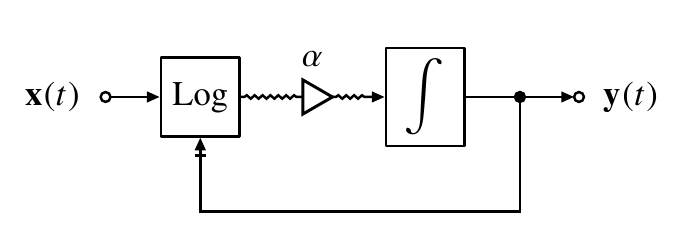}
  \caption[Geodesic one-pole block diagram]{The geodesic one-pole filter, given by
    \cref{eq:one-pole-manifold}, can be depicted as a block diagram.
    Each wire in this diagram represents a signal: solid wires
    represent signals whose values are hyperbolic points, and wavy
    wires represent signals whose values are hyperbolic vectors.  The
    triangular element represents multiplication.  The integral block
    represents the following relationship: the input to the block is
    the derivative of the block's output.  The arrow with a bar
    pointing to Log denotes the subscript argument.}
  \label{fig:one-pole-manifold-block}
\end{figure}
\begin{figure*}
  \subfloat[Geodesic one-pole filter]{%
  \includegraphics{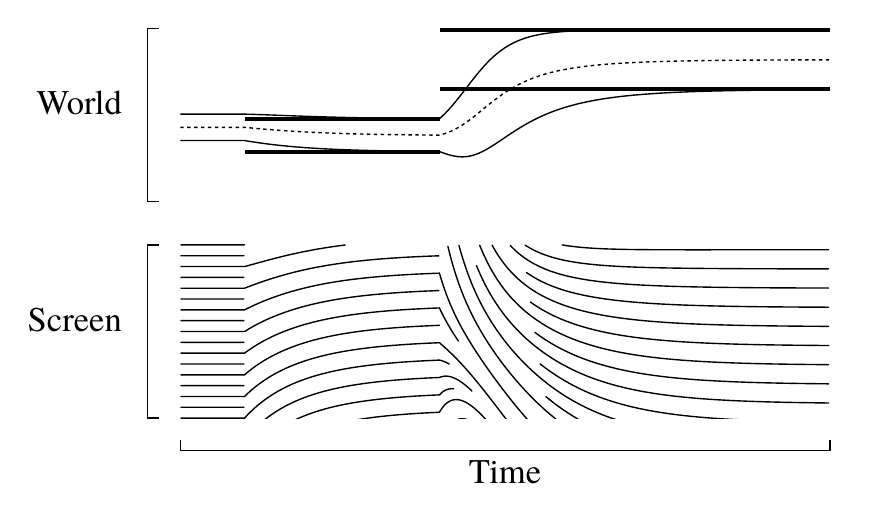}%
    \label{fig:one-pole-streamlines}}
  \subfloat[Clipped geodesic one-pole filter]{%
  \includegraphics{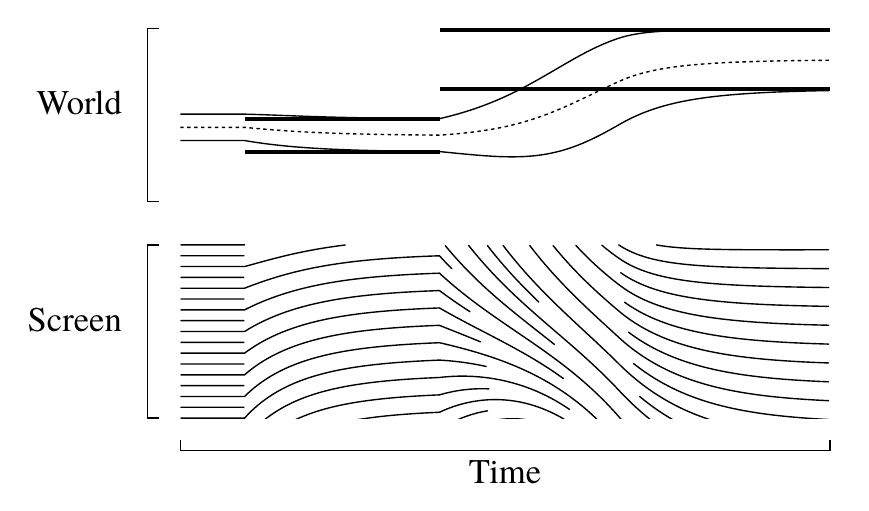}%
    \label{fig:clipped-one-pole-streamlines}}\\
  \subfloat[Cascaded one-pole filter]{%
  \includegraphics{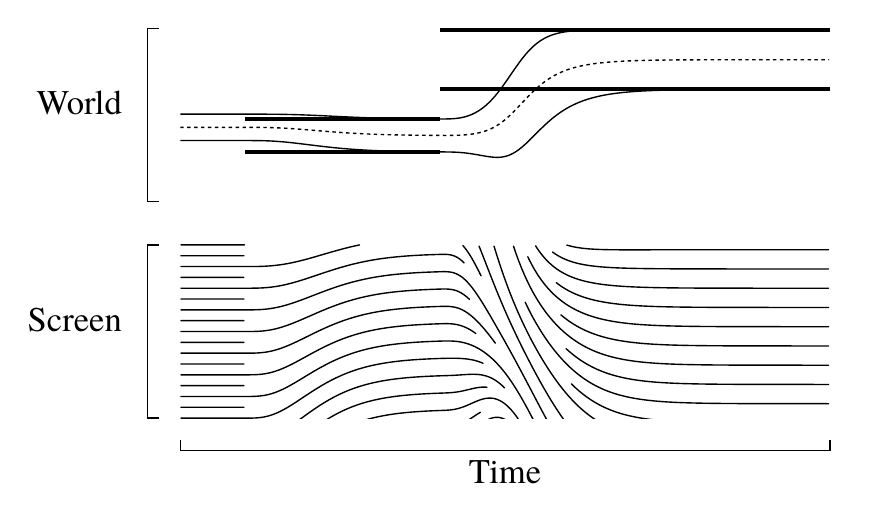}%
    \label{fig:four-pole-streamlines}}
  \subfloat[Clipped cascaded one-pole filter]{%
  \includegraphics{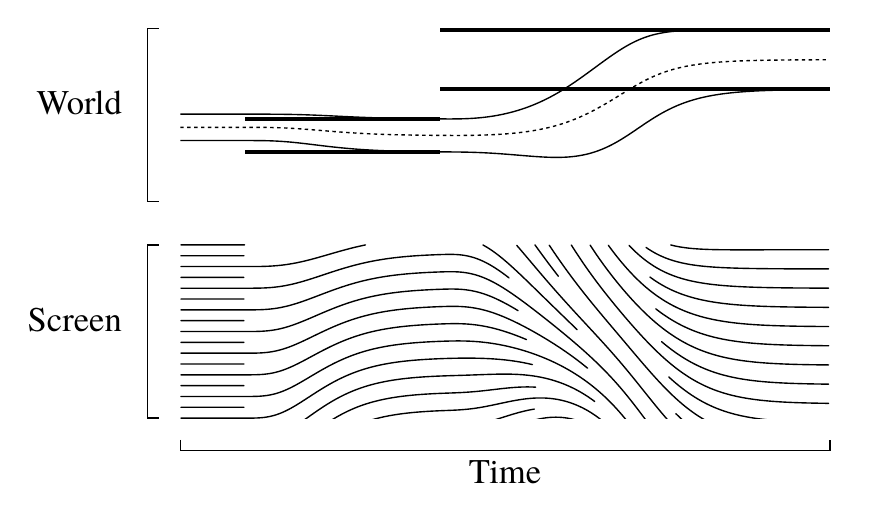}%
    \label{fig:clipped-four-pole-streamlines}}\\
  \caption[Geodesic one-pole vs clipped geodesic one-pole]{The geodesic one-pole
    filter (a) produces faster zooming and panning animations for targets that
    are farther away.  The clipped one-pole filter (b), by contrast, limits the
    maximum speed of the animation.  For discontinuous input, neither of these
    systems produces smooth zooming and panning animations, but they can be
    combined with additional one-pole filters to produce smoother animations, as
    shown in (c) and (d).}
  \label{fig:one-pole-vs-clipped}
\end{figure*}

For the target signal $\V x(t)$, the output $\V y(t)$ of the geodesic
one-pole filter is defined by the differential equation
\begin{equation}\label{eq:one-pole-manifold}
  \frac{d}{dt}\V y(t) = \alpha \Log_{\V y(t)} \V x(t)
\end{equation}
where $\alpha$ controls the speed of the animation.  This equation is
depicted in block diagram form in \cref{fig:one-pole-manifold-block}.
This differential equation can be explained as follows: at every
instant, walk along a geodesic towards the target with hyperbolic
speed proportional to the hyperbolic distance to the target.  This
system is clearly causal, since the derivative of the output depends
only on the current output and input.  The output of this system is
shown in \cref{fig:one-pole-streamlines}.

\subsection{Cascading filters}

While the one pole filter does smooth its input, it is not strong
enough for our purposes.  As can be seen from
\cref{eq:one-pole-manifold}, the derivative of the output is
proportional to the difference between the input and the output.
Because the output's derivative is always defined, the output will be
continuous even if the input is not.  However, it is clear from the
definition that a discontinuity in the input will introduce a velocity
discontinuity in the output.

\begin{figure}
  \centering
  \includegraphics{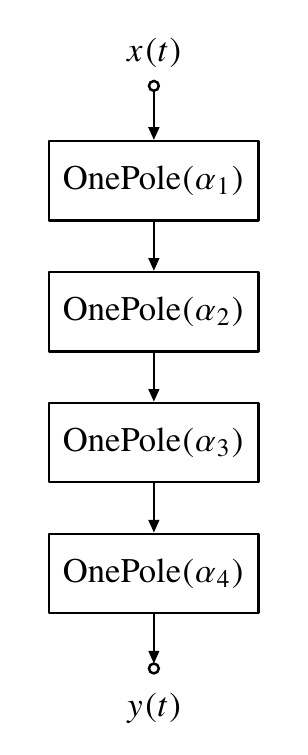}
  \caption[Cascaded one-pole filter block diagram]{The output of a single one-pole filter is continuous in
    position, but may not be continuous in velocity.  In order to
    guarantee that the output is smooth to a higher degree, one-pole
    filters can be chained together.  The filter depicted in this
    diagram, in which four one-pole filters are cascaded, produces an
    output with continuous position, velocity, acceleration, and jerk.
    Each block $\operatorname{OnePole}(\alpha)$ denotes the filter
    structure shown in \cref{fig:one-pole-manifold-block}.}
  \label{fig:cascaded-one-pole-block}
\end{figure}

To create a more powerful filter, we can cascade multiple instances of
the one pole filter, i.e. place them in series.  An example of this is
shown in \cref{fig:cascaded-one-pole-block}.  The output of the
cascaded one-pole filter is shown in
\cref{fig:cascaded-one-pole-hylerp}.  Each additional one-pole
filter used in a cascaded arrangement increases the order of
smoothness.  A single one-pole filter produces an output with
continuous position.  Cascading two one-pole filters produces an
output with continuous velocity.  Cascading three one-pole filters
produces an output with continuous acceleration.  Cascading four
one-pole filters produces an output with continuous jerk.

\subsection{Limiting the hyperbolic speed}

Like the easing curve method, the one-pole filter will produce
animations where the hyperbolic speed is greater for far-away targets
than for nearby targets.  For some use-cases, this can be undesirable,
and so we will now discuss a simple modification of the one-pole
filter that limits the maximum speed of the animation.

To accomplish this, we first define the function
$\operatorname{clipvec}(\V X, c)$ as follows:
\begin{equation}
  \operatorname{clipvec}(\V X, c)
  =
  \begin{cases}
  \V X & |\V X| < c\\
  c\frac{\V X}{|\V X|} & |\V X| \geq c
  \end{cases}
\end{equation}
This function takes a vector $\V X \in \mathbb T_{\V x}\mathbb H^n$ and modifies
its magnitude hyperbolic $|\V X|$ so that it does not exceed a threshold $c$.
Using this function, we can then define the clipped geodesic one-pole filter as
\begin{equation}\label{eq:clipped-one-pole}
  \frac{d}{dt}\V y(t)
  =
  \operatorname{clipvec}\Parens*{\alpha \Log_{\V y(t)} \V x(t), c}
\end{equation}
where $c$ is a threshold that sets the maximum allowed hyperbolic speed.  The
output of this system is shown in \cref{fig:clipped-one-pole-streamlines}.  By
placing in series a clipped one-pole filter followed by three geodesic one-pole
filters, a system is formed whose output is smooth and whose speed is limited.
We call this system the \emph{clipped cascaded geodesic one-pole filter}.  The
video demo that accompanies this paper uses the values $c = 1\text{ Hz}$ for the
first filter and $\alpha = 6\text{ Hz}$ for all four filters\footnote{These
  definitions of $c$ and $\alpha$ assume that distance in hyperbolic space is
  unitless, as is the case when the metric is unitless.  If the metric were
  defined strictly as the root-mean-square velocity of the optical flow on a
  physical screen, then $c$ would instead have units of speed (meters per
  second).  However, in this case, the value of $c$ would depend on the size of
  the screen.}.


\subsection{Discretizing the one-pole filter}

The previous section described solutions to the interruptible zooming
and panning problem in terms of cascaded first-order differential
equations.  To write a program that implements one of these solutions,
any ordinary differential equation (ODE) solver can be used to
numerically approximate these differential equations.  In this
section, however, we take an alternative approach: we convert the
continuous-time systems described in the last section into
discrete-time systems.  The solution we arrive at is quite simple, and
does not require the use of an external ODE solver.

Notice that the geodesic one-pole filter \cref{eq:one-pole-manifold}
and the clipped geodesic one-pole filter \cref{eq:clipped-one-pole}
have the form
\begin{equation}\label{eq:first-order-diffeq}
  \frac{d}{dt}\V y(t) = f(\V y(t), \V x(t))
\end{equation}
where $f(\V y(t), \V x(t)) \in \mathbb T_{\V y(t)}$.  A differential
equation of this form can be approximated by
\begin{equation}\label{eq:first-order-approx}
  \V y[i] = \Exp(T f(\V y[i-1], \V x[i]))
\end{equation}
where $T$ is the sampling period and $f(\V y[i-1], \V x[i]) \in
\mathbb T_{\V y[i-1]}$.  To see that this does approximate the
continuous-time system, note that \cref{eq:first-order-approx} can be
written as
\[
  \V y(t) = \Exp(T f(\V y(t-T), \V x(t))).
\]
Taking the logarithmic map of both sides with base $\V y(t - T)$ and
then dividing both sides by $T$ gives
\[
  \frac{\Log_{\V y(t - T)}\V y(t)}{T} = f(\V y(t - T), \V x(t))
\]
Taking the limit as $T$ goes to zero gives
\cref{eq:first-order-diffeq}.

\begin{figure}
  \centering
  \vspace{-0.8em}
  \includegraphics{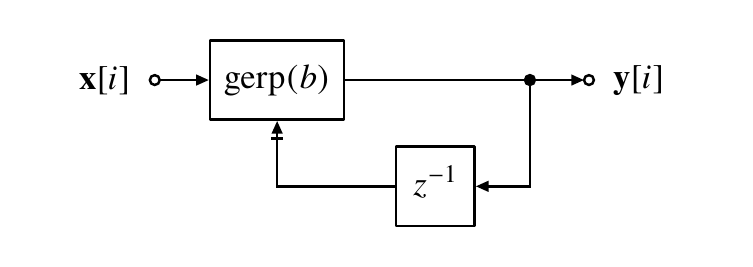}
  \caption[Discrete-time geodesic one-pole filter]{The discrete-time hyperbolic one-pole filter is an
    approximation to the continuous-time one-pole filter depicted in
    \cref{fig:one-pole-manifold-block}.  The element $z^{-1}$ denotes
    a delay by one sample, i.e. a delay by the sampling period $T$.
    The block $\operatorname{gerp}(b)$ denotes $\operatorname{gerp}(\V
    x_0, \V x_1, b)$, where the arrowhead with a bar in the diagram
    corresponds to $\V x_0$ and the arrowhead without a bar in the
    diagram corresponds to $\V x_1$.}
  \label{fig:discrete-one-pole-block}
\end{figure}

Applying this approximation to the geodesic one-pole filter
\cref{eq:one-pole-manifold} and simplifying using
\cref{eq:gerp-exp-log} gives
\begin{equation}\label{eq:discrete-one-pole}
  \V y[i] = \operatorname{gerp}(\V y[i-1],\V x[i],b)
\end{equation}
where $b = \alpha T$ for sampling period $T$.  In other words, at each
frame, we take step towards the target, where the length of the step
is proportional to the distance to the target.  A block diagram of
this approximation is shown in \cref{fig:discrete-one-pole-block}.
Note that this approximation oscillates for $b > 1$ and is unstable
for $b > 2$, and so the sampling period $T$ should be small enough
that $b < 1$.

\begin{figure}
  \centering
  \includegraphics{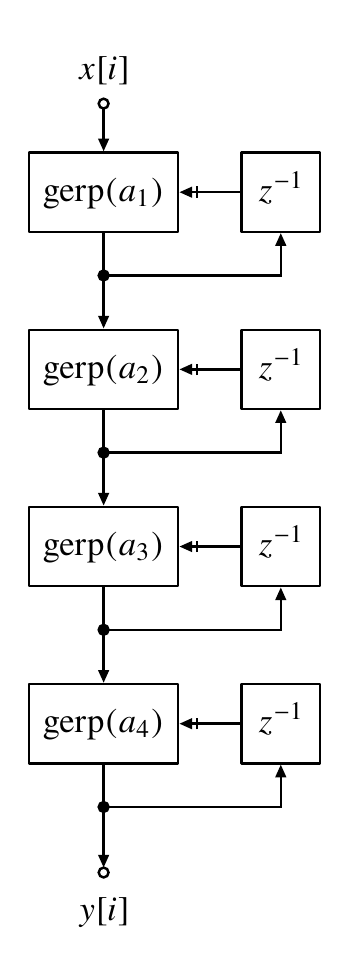}
  \caption[Discretized cascaded one-pole filter]{The cascaded one-pole system depicted in
    \cref{fig:cascaded-one-pole-block} can be discretized by cascading
    the discrete one-one pole blocks depicted in
    \cref{fig:discrete-one-pole-block}.}
  \label{fig:cascaded-discrete-one-pole-block}
\end{figure}

To approximate the cascaded one-pole system shown in
\cref{fig:cascaded-one-pole-block}, we can simply cascade the discrete
one-pole filters as shown in
\cref{fig:cascaded-discrete-one-pole-block}.

Applying the approximation given by
\cref{eq:first-order-diffeq,eq:first-order-approx} to the clipped
one-pole filter \cref{eq:clipped-one-pole} and simplifying gives
\begin{subequations}
\begin{align}
  \V y[i] &= \operatorname{geo}\Parens*{\V y[i-1], \V x[i], s[i]}\\
  s[i] &= T \min\Parens*{c, \alpha S[i]}\\
  S[i] &= \operatorname{dist}(\V y[i-1], \V x[i])
\end{align}
\end{subequations}
where $T$ is the sampling period.






\section{Smoothing using geodesic two-pole filters}

The technique discussed in the previous section formed higher-order
filters by cascading one-pole filters.  Animations that make
use of these filters travel directly to their target, settling in
slowly.  Suppose that we instead wanted an animation where the camera
overshoots the target slightly before settling in a spring-like
motion.  To achieve this effect, we can use a two pole filter, which
models the behavior of a spring-mass-damper system.

\begin{figure}
  \centering
  \includegraphics{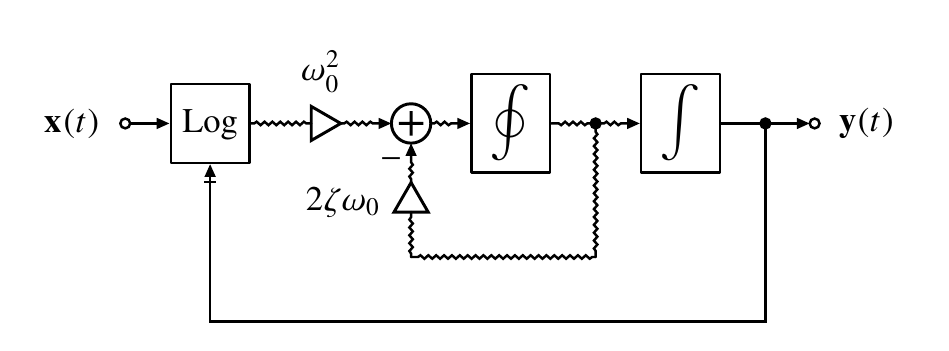}
  \caption[Geodesic two-pole filter block diagram]{The geodesic two-pole filter defined by
    \cref{eq:manifold-two-pole} can be depicted as a block diagram.
    Here, the integral with a circle represents the following
    relationship: the input to the block is the covariant derivative
    of the output.}
  \label{fig:manifold-two-pole-block}
\end{figure}

\begin{figure}
  \centering
  \includegraphics{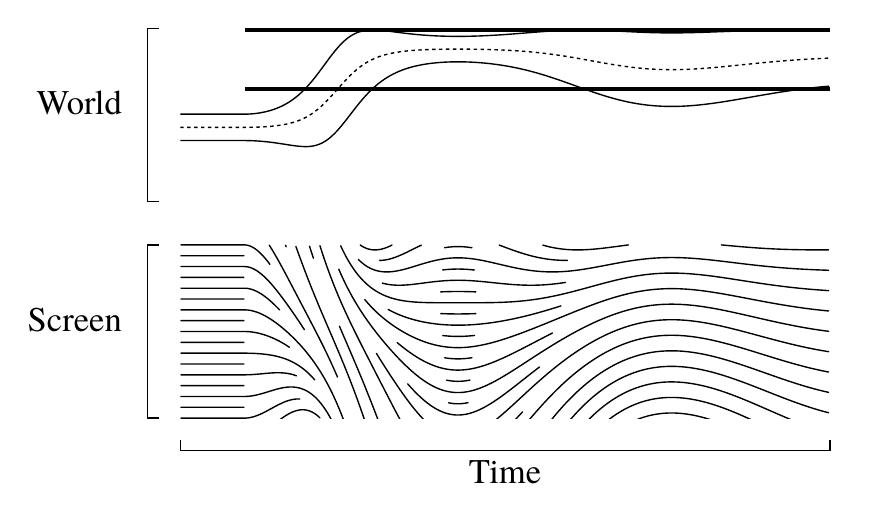}
  \caption[Geodesic one-pole filter results]{The geodesic two-pole filter defined by
    \cref{eq:manifold-two-pole} produces spring-like motion.}
  \label{fig:two-pole-streamlines}
\end{figure}

In Euclidean space, a spring-mass-damper system experiences two
forces: a spring force that pulls the mass towards the equilibrium
position and is proportional to the displacement from this equilibrium
position, and a damping force that pushes against the mass's
velocity.  The spring-mass-damper system can be adapted to hyperbolic
space as follows:
\begin{equation}\label{eq:manifold-two-pole}
  \frac{D}{dt}\frac{d}{dt}\V y(t)
  = \omega_0^2\Log_{\V y(t)}(\V x(t)) - 2\zeta\omega_0\frac{d}{dt}\V y(t)
\end{equation}
The variable $\omega_0$ controls the rate of the animation, and the
parameter $\zeta$ controls the damping.  When $\zeta < 1$, the system
oscillates in a spring-like motion as it settles.  A block diagram of
this is shown in \cref{fig:manifold-two-pole-block}.  The response of
the two-pole filter for discontinuous input is shown in
\cref{fig:two-pole-streamlines}.

The system defined by \cref{eq:manifold-two-pole} can of course be
placed in series with a one-pole filter or another two-pole filter.
Like the one-pole filter and easing curves, animation produced by the
two-pole filter will be faster for distant targets than for nearby
targets.  A simple way to address this is to place the clipped
one-pole filter in series with the two-pole filter.  We call the
system consisting of a single clipped one-pole filter followed by a
two-pole filter the \emph{clipped geodesic two-pole filter}.  Another
way to reduce the speed for distant targets would be to use quadratic
damping instead of linear damping, but this paper does not explore
this option.

\subsection{Discretizing the two-pole filter}

As with the continuous-time one-pole filter, the differential equation
given by \cref{eq:manifold-two-pole} can be approximated using a
standard ODE solver.  However, we can also write a simple
approximation by converting the continuous-time system to a
discrete-time system.

Notice that the geodesic two pole filter has the form
\begin{equation}
  \frac{D}{dt}\frac{d}{dt}\V y(t)
  = f\Parens*{\V y(t), \V{\dot y}(t), \V x(t)}
\end{equation}
where $\V{\dot y}(t)$ denotes the time derivative of $\V y(t)$.  A
system of this form can be approximated by
\begin{align}
  \V{\dot y}[i] &= \V{\dot y}[i-1] + Tf\Parens*{\V y[i-1],\,\TrMap_{\V y[i-1]}\V{\dot y}[i-1],\,\V x[i]}\\
  \V{y}[i] &= \Exp(T\V{\dot y}[i])
\end{align}
Here both $\V y[i] \in \mathbb H^n$ and $\V{\dot y}[i] \in \mathbb
T_{\V y[i-1]}\mathbb H^n$ are state variables that are updated at each
frame.  The parallel transport in this formula is necessary to move
the vector variable $\V{\dot y}[i]$ to the correct location each time
the position variable $\V y[i]$ is updated.

Applying this approximation to the geodesic two-pole filter gives
\begin{align}
  \V{\dot y}[i] &= (1 - 2T\zeta\omega_0)\TrMap_{\V y[i-1]}(\V{\dot y}[i-1])\\
  &\quad{}
                  + T\omega_0^2\Log_{\V y[i-1]}(\V x[i])\notag\\
  \V{y}[i] &= \Exp(T\V{\dot y}[i]).
\end{align}
The demo videos that accompany this paper have sixty frames per second, and so
$T = (1/60)\text{ s}$.

\section{Conclusion and future work}

We have introduced the hyperbolic model of zooming and panning.  This
model exploits the Poincar\'e half-space model of hyperbolic space to
model zooming and panning.  Using this model, we have introduced novel
zooming and panning animation techniques based on signal processing
and Riemannian geometry.  These animation techniques produce smooth
zooming and panning animations to a target value, and these animations
remain smooth even when the animation is interrupted by a change of
target.  Such interruptions are common in interactive uses, where user
input can interrupt zooming and panning animations at any time.

We have also introduced world\slash{}screen diagrams to visualize
zooming and panning animations.  These diagrams emphasize the
perceptual aspects of zooming and panning, and are useful for
evaluating different zooming and panning techniques.

There are many avenues for future work.  For example, if instead of the
interactive zooming and panning problem we have investigated in this paper, we
merely wish to interpolate a set of target points, the B\'ezier curve
generalization of Park and Ravani \cite{ravani1995bezier} can be
straightforwardly applied to zooming and panning using the definitions we
presented here.  The animation model in this paper could also be applied to
other navigation problems, e.g. lenses \cite{appert2010high,bier1993toolglass}
or zooming and panning on a globe.  Finally, there are possible variations on
the specific signal processing systems we have introduced that could be explored
in future work.

This paper has used the Poincar\'e half-space model to represent
zooming and panning positions.  However, there are many other possible
coordinate systems that could be used for hyperbolic space.  While the
results remain the same regardless of coordinate system, different
coordinate systems might be more convenient for particular purposes.
Other well-known representations for hyperbolic space include the
Poicar\'e disk model, the Klein model, and the hyperboloid model
\cite{cannon1997hyperbolic}.  Another option would be to use
space-scale diagrams.  Yet another option would be to a model in which
the altitude of the camera is constant and the angle of view is varied
to zoom and pan.  Such a model would match the term ``zooming'' more
closely, since zooming on a camera involves changing the angle of view
rather than changing the position of the camera.\footnote{In fact, it
  might be more accurate to refer to the model used in this paper as
  dollying and panning rather than zooming and panning.}

\section{Acknowledgments}

This research was supported in part by NSF grant IIS-1447416.

%% file: hyperbolic-dynamics.bbl
\begin{thebibliography}{10}

\bibitem{appert2010high}
C.~Appert, O.~Chapuis, and E.~Pietriga.
\newblock High-precision magnification lenses.
\newblock In {\em Proceedings of the SIGCHI Conference on Human Factors in
  Computing Systems}, pages 273--282. ACM, 2010.

\bibitem{bederson2000jazz}
B.~B. Bederson, J.~Meyer, and L.~Good.
\newblock Jazz: an extensible zoomable user interface graphics toolkit in java.
\newblock In {\em Proceedings of the 13th annual ACM symposium on User
  interface software and technology}, pages 171--180. ACM, 2000.

\bibitem{bier1993toolglass}
E.~A. Bier, M.~C. Stone, K.~Pier, W.~Buxton, and T.~D. DeRose.
\newblock Toolglass and magic lenses: the see-through interface.
\newblock In {\em Proceedings of the 20th annual conference on Computer
  graphics and interactive techniques}, pages 73--80. ACM, 1993.

\bibitem{cannon1997hyperbolic}
J.~W. Cannon, W.~J. Floyd, R.~Kenyon, W.~R. Parry, et~al.
\newblock Hyperbolic geometry.
\newblock {\em Flavors of geometry}, 31:59--115, 1997.

\bibitem{cockburn2008review}
A.~Cockburn, A.~K. Karlson, and B.~B. Bederson.
\newblock A review of overview+ detail, zooming, and focus+ context interfaces.
\newblock {\em ACM Comput. Surv.}, 41(1):2--1, 2008.

\bibitem{do1992riemannian}
M.~P. Do~Carmo and J.~Flaherty~Francis.
\newblock {\em Riemannian geometry}, volume 115.
\newblock Birkh{\"a}user Boston, 1992.

\bibitem{dragicevic2011temporal}
P.~Dragicevic, A.~Bezerianos, W.~Javed, N.~Elmqvist, and J.-D. Fekete.
\newblock Temporal distortion for animated transitions.
\newblock In {\em Proceedings of the SIGCHI Conference on Human Factors in
  Computing Systems}, pages 2009--2018. ACM, 2011.

\bibitem{furnas1995space}
G.~W. Furnas and B.~B. Bederson.
\newblock Space-scale diagrams: Understanding multiscale interfaces.
\newblock In {\em Proceedings of the SIGCHI conference on Human factors in
  computing systems}, pages 234--241. ACM Press/Addison-Wesley Publishing Co.,
  1995.

\bibitem{guiard2004view}
Y.~Guiard, M.~Beaudouin-Lafon, J.~Bastin, D.~Pasveer, and S.~Zhai.
\newblock View size and pointing difficulty in multi-scale navigation.
\newblock In {\em Proceedings of the working conference on Advanced visual
  interfaces}, pages 117--124. ACM, 2004.

\bibitem{igarashi2000speed}
T.~Igarashi and K.~Hinckley.
\newblock Speed-dependent automatic zooming for browsing large documents.
\newblock In {\em Proceedings of the 13th annual ACM symposium on User
  interface software and technology}, pages 139--148. ACM, 2000.

\bibitem{lee1997riemannian}
J.~Lee.
\newblock {\em Riemannian manifolds : an introduction to curvature}.
\newblock Springer, New York, 1997.

\bibitem{lee2012introduction}
J.~Lee.
\newblock {\em Introduction to smooth manifolds}.
\newblock Springer, New York London, 2012.

\bibitem{lee2002general}
J.~Lee and S.~Y. Shin.
\newblock General construction of time-domain filters for orientation data.
\newblock {\em Visualization and Computer Graphics, IEEE Transactions on},
  8(2):119--128, 2002.

\bibitem{matvienko2013dense}
V.~Matvienko and J.~Kr{\"u}ger.
\newblock Dense isocontour imaging.
\newblock In {\em SIGGRAPH Asia 2013 Technical Briefs}, page~16. ACM, 2013.

\bibitem{osborne2013geodesic}
J.~M. Osborne and G.~P. Hicks.
\newblock The geodesic spring on the euclidean sphere with
  parallel-transport-based damping.
\newblock {\em Notices of the American Mathematical Society}, 60(5):544--557,
  2013.

\bibitem{ravani1995bezier}
B.~Ravani and F.~C. Park.
\newblock B\'ezier curves on riemannian manifolds and lie groups with
  kinematics applications.
\newblock 1995.

\bibitem{reach2017signals}
A.~M. Reach and C.~North.
\newblock The signals and systems approach to animation, 2017.

\bibitem{van2003smooth}
J.~J. van Wijk and W.~A. Nuij.
\newblock Smooth and efficient zooming and panning.
\newblock In {\em Information Visualization, 2003. INFOVIS 2003. IEEE Symposium
  on}, pages 15--23. IEEE, 2003.

\bibitem{van2004model}
J.~J. Van~Wijk and W.~A. Nuij.
\newblock A model for smooth viewing and navigation of large 2{D} information
  spaces.
\newblock {\em IEEE Transactions on Visualization and Computer Graphics},
  10(4):447--458, 2004.

\end{thebibliography}
